\documentclass[prb,floatfix,superscriptaddress,twocolumn,aps,10]{revtex4-2}
\usepackage[english]{babel}
\usepackage{bm}
\usepackage{graphicx}
\usepackage{amsmath}
\usepackage{amssymb}
\usepackage{wasysym}
\usepackage{comment}
\usepackage[dvipsnames]{xcolor}
\usepackage{tikz}
\usepackage{adjustbox}
\usepackage{booktabs}
\usepackage{array}
\usepackage{multirow}
\usepackage{parskip,physics}
\usepackage[inkscapelatex=false]{svg}
\usepackage{subcaption}
\usepackage{ragged2e}

\usepackage[hidelinks]{hyperref}

\begin{document}

\title{Exact ground state on the 3D analogue of the Shastry-Sutherland model}
    \author{Kelvin Salou-Smith}
	\email[]{kelvin.salou-smith@u-bordeaux.fr}
    \affiliation{CNRS, Universit\'e de Bordeaux, LOMA, UMR 5798, 33400 Talence, France}
		 
	\author{Arnaud Ralko}
	\email[]{arnaud.ralko@neel.cnrs.fr}
	\affiliation{Institut N\'eel, UPR2940, Universit\'e Grenoble Alpes et CNRS, Grenoble FR-38042, France}

    \author{Ludovic D.C. Jaubert}
	\email[]{ludovic.jaubert@u-bordeaux.fr}
    \affiliation{CNRS, Universit\'e de Bordeaux, LOMA, UMR 5798, 33400 Talence, France}
		
	\date{\today}
		
\begin{abstract} 
Exact results in frustrated quantum many-body systems are rare, especially in dimensions higher than one. The Shastry-Sutherland (SS) model stands out as a rare example of a two-dimensional spin system with an exactly solvable dimer singlet ground state. In this work, we introduce a three-dimensional analogue of the SS lattice, constructed by deforming the pyrochlore lattice to preserve the local SS geometry. Despite the dimensional increase and altered topology, the ground-state phase diagrams of classical Ising and Heisenberg spins, remain analytically tractable and closely follow their 2D counterparts, including the existence of a $1/3$ magnetization plateau and umbrella states. Most notably, for quantum spins $S=1/2$, the dimer singlet state survives as an exact ground state over a finite region of the phase diagram. We argue, using exact diagonalization, that the singlet phase is stabilized beyond its 2D counterpart, suggesting enhanced robustness in three dimensions. These results offer a rare, controlled platform to explore the impact of dimensionality on quantum frustration, exact solvability, and potential spin liquid behavior in 3D, with relevance to emergent topological and magnetic phases.
\end{abstract}
\maketitle

Exact results are more the exception than the norm in frustrated quantum many-body physics. While there are well-known examples in (quasi-)one dimension (1D) such as the sawtooth chain and Majumdar-Ghosh model \cite{Miyahara2011}, they tend to become scarcer as the system dimension increases. In 2D, a few models stand out for admitting exact solutions under specific geometric or algebraic constraints. The Kitaev honeycomb model \cite{Kitaev06a} and the Toric code \cite{Kitaev03a} are famous for realizing quantum spin liquids with emergent Majorana fermions and nontrivial topological order. The Shastry-Sutherland and maple-leaf lattices \cite{shastry1981,ghosh2022} host a dimer ground state formed as a tensor product of localized singlets, while the Rokhsar-Kivelson quantum dimer model \cite{Rokhsar88a} features a ground state built from highly resonating superpositions of short-range valence bond configurations.

Among these, the Shastry-Sutherland model has been actively studied since its experimental realization in SrCu$_2$(BO$_3$)$_2$ \cite{kageyama1999}, particularly for its rich sequence of magnetization plateaux \cite{onizuka2000,corboz2014} and, more recently, as a quantum magnetic analogue of a critical point \cite{jimenez2021}. Its phase diagram has also been proposed to host deconfined quantum criticality \cite{lee2019}, pushed to finite magnetic field above a probable quantum spin liquid ground state sandwiched between a plaquette valence bond solid and Néel order at low field \cite{yang2022}, as recently explored using neural quantum states \cite{mezera2023,viteritti2025,nutakki2025}. 


\begin{figure}
    \centering
    \includegraphics[width=0.9\columnwidth]{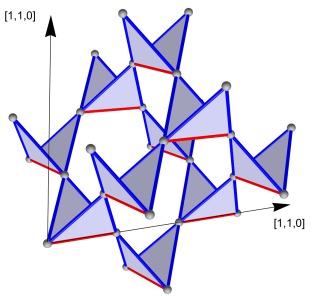}
    \caption{\justifying\small
    The 3D Shastry-Sutherland lattice, which is locally indistinguishable from its 2D counterpart. Each four-site unit cell is made of four $J_1$ and one $J_2$ coupling (thin and thick bonds respectively). Beyond the four-site unit cell, the smallest loop is made of 7 spins. The structure is equivalent to the pyrochlore lattice where bonds at the top of all tetrahedra have been removed.
    }
    \label{fig:latt}
\end{figure}

SrCu$_2$(BO$_3$)$_2$ is actually a stack of Shastry-Sutherland (SS) lattices \cite{ueda1999}, where it was quickly realized that the dimer singlet phase of the 2D model remains the ground state for part of the phase diagram with weak inter-layer coupling \cite{koga2000,miyahara2000,carpentier2001}. This has led to a series of 3D parent Hamiltonians with an exact quantum dimer singlet ground state, generally based on the cubic lattice with appropriate further-neighbor couplings \cite{chen2002,surendran2002,shik2003,sun2021} or on (bi-)pyramidal units with a square basis \cite{kumar2002}.

In this paper, we introduce the three-dimensional analogue of the 2D Shastry-Sutherland lattice, in the sense that they are locally indistinguishable [Fig.~\ref{fig:latt}]. The 3D structure is paved by corner-sharing four-site unit cells, each unit cell forming a square of $J_1$ bonds and one diagonal $J_2$ coupling, and each site having a connectivity of five. After introducing the lattice and the Hamiltonian, we will first solve the exact classical ground-state phase diagram with Ising and Heisenberg spins. We show that the phase diagram is the same as in 2D, with the caveat that the $1/3$ magnetization (Ising) and umbrella states (Heisenberg) need to be accommodated on the 3D structure. Then, we consider quantum spins $S=1/2$ whose ground state can be derived exactly for part of the phase diagram, following the reasoning of Shastry \& Sutherland \cite{shastry1981,sutherland1983}. Exact Diagonalization (ED) suggests an enhanced stability of the dimer singlet phase in 3D.

\section{Model}
\label{sec:model}

The two-dimensional SS lattice can be viewed as a modified checkerboard, where one diagonal bond is removed from each square plaquette. This leads to a uniform coordination number of five per site, comprising four nearest-neighbor $J_1$ bonds and a single diagonal $J_2$ coupling.\\
We extend this construction to three dimensions by starting from the pyrochlore lattice, which consists of corner-sharing tetrahedra. Selecting the $z$-axis as a reference, we systematically remove the “top” bond from each tetrahedron and designate the opposite “bottom” bond as the $J_2$ interaction [Fig.~\ref{fig:latt}]. This results in a structure that is locally indistinguishable from the 2D SS lattice. In particular, the $J_1$ bonds form a bipartite network capable of supporting Néel order at large $J_1$, while every site participates in exactly one $J_2$ bond. The resulting spin Hamiltonian is:
\begin{align}
    \mathcal{H} = J_1 \sum_{\langle i,j \rangle}  \mathbf{S}_i \cdot \mathbf{S}_j + J_2 \sum_{\langle\langle i,j \rangle\rangle}  \mathbf{S}_i \cdot \mathbf{S}_j - \mathbf{h}\cdot \sum_{i} \mathbf{S}_i,
    \label{eq:ham}
\end{align}
where $\mathbf{S}_i$ is the magnetic moment at site $i$ and $\mathbf{h}=h \,\mathbf{e}_z$ is along the vertical axis. From now on, we set the energy scale with $J_1=1$ and will only consider antiferromagnetic $J_2>0$.

There is, however, a crucial difference between the 2D and 3D Shastry-Sutherland lattices. In two dimensions, the smallest closed loop beyond the unit cell consists of four $J_1$ bonds. In three dimensions, the minimal loop becomes significantly longer, comprising seven bonds: six $J_1$ and one $J_2$ [Fig.~\ref{fig:latt}]. As a result, many properties derived in 2D remain valid in 3D, but with some modifications. In the following sections, we explore these differences systematically by considering, in turn, classical Ising spins, classical Heisenberg spins, and quantum spins with $S=1/2$.

\section{Classical Ising spins}

Here we consider $\mathbf{S}_i=\sigma_i\,\mathbf{e}_z$ with $\sigma_i=\pm 1$. The ground state phase diagram at zero field is trivial and independent of the system dimension, with N\'eel order for $J_2<2 J_1$ and the well-known Ising dimer phase for $J_2>2 J_1$ ($\uparrow\downarrow$ or $\downarrow\uparrow$ on all $J_2$ bonds). But in a field $h$, the phase diagram is somewhat more subtle to derive exactly.

This was done by Dublenych in 2D \cite{Dublenych2012}, who showed that the $1/3$ magnetization plateau was the only possible fractional plateau. Dublenych's method relies on determining which set of triangular spin states are expected in the ground state for a given region of the phase diagram, based on a form of Maxwell construction in a multi-dimensional parameter space. Since our 3D SS lattice has been designed to reproduce the 2D one locally, the triangular spin states have the same energy and the method of Ref.~\cite{Dublenych2012} applies directly to our 3D model \cite{SM}. The only difference is about how to pave the lattice with these triangular states. The paving is straightforward for the N\'eel, dimer and ferromagnetic phases, but the absence of a four-site loop in 3D means that the 1/3 plateau needs to be revisited. Arbitrarily choosing $h>0$, the only triangular states allowed in this region of the phase diagram are \cite{Dublenych2012}:

\begin{center}
\begin{tikzpicture}
\begin{scope}[xshift=0cm]
\draw[line width=2pt] (0,0.3)--(0.3,0);
\filldraw[fill=black] (0,0)   circle (3pt);
\filldraw[fill=black] (0,0.3) circle (3pt);
\filldraw[fill=white] (0.3,0) circle (3pt);
\end{scope}
\begin{scope}[xshift=1cm]
\draw[line width=2pt] (0,0.3)--(0.3,0);
\filldraw[fill=black] (0,0)   circle (3pt);
\filldraw[fill=white] (0,0.3) circle (3pt);
\filldraw[fill=black] (0.3,0) circle (3pt);
\end{scope}
\begin{scope}[xshift=2cm]
\draw[line width=2pt] (0,0.3)--(0.3,0);
\filldraw[fill=white] (0,0)   circle (3pt);
\filldraw[fill=black] (0,0.3) circle (3pt);
\filldraw[fill=black] (0.3,0) circle (3pt);
\end{scope}
\end{tikzpicture}
\end{center}

\begin{figure}
    \centering
    \begin{subfigure}[b]{1.0\columnwidth}
        \includegraphics[width=0.8\columnwidth]{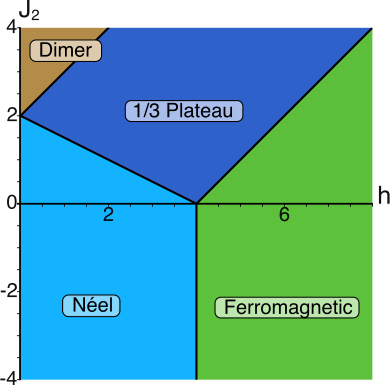}
        \caption{}
    \end{subfigure}
    \begin{subfigure}[b]{1.0\columnwidth}
        \includegraphics[width=0.8\columnwidth]{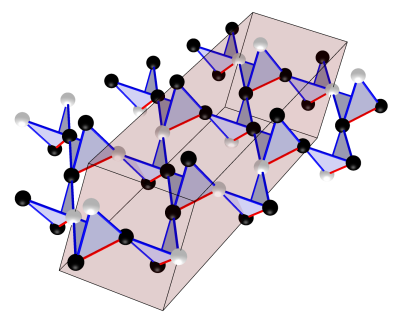}
        \caption{}
    \end{subfigure}
    \caption{\justifying\small
    \textit{Ising model:} (a) The ground-state phase diagram on the Shastry-Sutherland lattice is the same in 2D as in 3D. (b) The $1/3$ plateau needs, however, to be adapted to the three-dimensional geometry.
    }
    \label{fig:Ising}
\end{figure}

In these configurations, $\circ$ and $\bullet$ are respectively down and up spins, and the diagonal link is the $J_2$ coupling. Fortunately, it is possible to pave the entire 3D lattice with these three triangular states only, as illustrated in Fig.~\ref{fig:Ising}(b). This state also forms a $1/3$ magnetization plateau with a magnetic unit cell of 12 sites. As a consequence, we recover the same ground-state phase diagram in 3D as in 2D [Fig.~\ref{fig:Ising}(a)].

\section{Classical Heisenberg spins}

Now we consider $\mathbf{S}_{i}$ as a classical Heisenberg spins on a sphere of unit radius. Each triangle $t$ has three spins $\{0,1,2\}$, the latter two being linked by $J_2$. Let us define $\mathbf{X}_t = \mathbf{S}_{t,0} + J_2\left(\frac{\mathbf{S}_{t,1} +\mathbf{S}_{t,2}}{2}\right)$ \cite{Bilitewski2017} for both the planar ($\perp$) and orthogonal ($z$) spin components. Hamiltonian (\ref{eq:ham}) can be rewritten as (up to a constant)
\begin{align}
    \mathcal{H} = \sum_{t} \frac{\mathbf{X}_{\perp,t}^2}{J_2} + \frac{X_{z,t}^2}{J_2} - \frac{h \cdot X_{z,t}}{1 + J_2}.
\end{align}
There are two cases to distinguish \cite{Moliner2009,Grechnev2013} as outlined below:

\textbf{Case 1, $J_2\geq 1$ :} 
It is possible to impose $\mathbf{X}_{\perp,t}=0$ in order to minimize the orthogonal part of the energy. If we define a local $x$ axis which aligns with $\mathbf{S}_{\perp, t,0}$, then we have $S_{x, t, 1} = S_{x, t, 2} = - \frac{S_{x, t, 0}}{J_2}$ and $S_{y, t, 1} = -S_{y, t, 2}$. Defining the angle between $\mathbf{S}_{\perp, t,0}$ and $\mathbf{S}_{\perp, t,1}$ (resp. $\mathbf{S}_{\perp, t,2}$) as $\pi-\theta$ (resp. $\pi+\theta$), we have $\cos(\theta)=\frac{1}{J_2}$. Keeping in mind that each spin belongs to three triangles where it takes alternatively values 0, 1 and 2 [Fig.~\ref{fig:latt}], and in order to pave the entire lattice, we take the absolute value of the planar spin components to be equal. This also imposes the absolute value of the $z$ spin component to be equal. Since $\mathbf{X}_{\perp,t} = 0$, we can minimize the total energy of each triangle $t$ with $X_{z, t} = \frac{h\, J_2}{2(1+J_2)}$. It gives $S_{z, t,0} = S_{z, t,1} = S_{z, t,2} = \frac{h\, J_2}{2(1+J_2)^2}$. Therefore the saturation field, where all spins are aligned with the external field, is $h_{sat} = \frac{2(1+J_2)^2}{J_2}$.\\

\textbf{Case 2, $J_2\leq 1$ :} 
It is not possible to make $\mathbf{X}_{\perp,t}$ vanish anymore. However, $\mathbf{X}_{\perp,t}$ can be minimized by taking $\mathbf{S}_{\perp,t,1} = \mathbf{S}_{\perp,t,2} = -\mathbf{S}_{\perp,t,0}$ such that $\mathbf{X}_{\perp,t} = (1 - J_2)\mathbf{S}_{\perp,t,0}$. By defining the angle between the spin direction and the $x\text{-}y$ plane as $\phi$, we can write the energy of a triangle as 
\begin{align}
    \mathcal{H}_t(\phi) = \frac{1 + J_2^2}{J_2} + 2(2\sin^2{\phi} - 1) - h\sin{\phi},
\end{align}
which is minimal for $\phi = \arcsin{(h/8)}$. The system is thus in a N\'eel state, canted in the direction of the external field until a saturation field strength of $h_{sat} = 8$.\\

\begin{figure}
    \centering
    \begin{subfigure}[b]{1.0\columnwidth}
        \includegraphics[width=0.8\columnwidth]{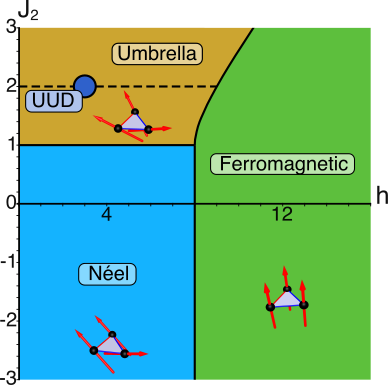}
        \caption{}
    \end{subfigure}
    \begin{subfigure}[b]{1.0\columnwidth}
        \includegraphics[width=0.8\columnwidth]{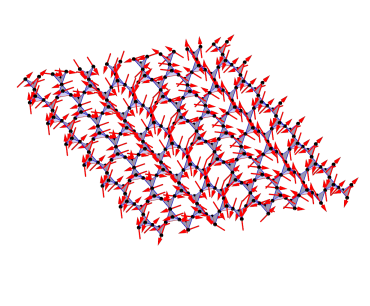}
        \caption{}
    \end{subfigure}
    \caption{\justifying\small
    \textit{Heisenberg model:} (a) The ground-state phase diagram on the Shastry-Sutherland lattice is the same in 2D as in 3D. The dashed line indicates the 120$^\circ$ phase. (b) The spiral state for $J_2\geq 1$ needs, however, to be adapted to the three-dimensional geometry.
    }
    \label{fig:Heis}
\end{figure}

These results are the same as in 2D \cite{Moliner2009,Grechnev2013}. The question is now whether these local constraints can be satisfied on the 3D SS lattice. For the latter case, this is automatic since it simply requires a tilted N\'eel state. While less obvious, this is also possible for $J_2\geq 1$, as illustrated in Fig.~\ref{fig:Heis}.(b). It is possible to close around the seven-site loop while accommodating the local spin rotation within each triangle. We thus recover the same ground-state phase diagram as in 2D [Fig.~\ref{fig:Heis}.(a)] \cite{Moliner2009,Grechnev2013}, albeit with an umbrella state that propagates in three dimensions. In particular, the up-up-down collinear state is also part of the ground state at $J_2=2$ and $h=3$, which will likely be favored entropically at finite temperatures leading to a $1/3$ magnetization plateau\cite{Moliner2009}.


\section{Quantum spins $S=1/2$}

When considering quantum spins$-1/2$ in zero field, it is best to rewrite the Hamiltonian as
\begin{align}
    \mathcal{H} = \sum_{\mathrm{t}} (\mathbf{\hat S}_{t,0} \cdot \mathbf{\hat S}_{t,1} + \mathbf{\hat S}_{t,0} \cdot \mathbf{\hat S}_{t,2}) + \frac{J_2}{2} \, \mathbf{\hat S}_{t,1} \cdot \mathbf{\hat S}_{t,2}.
    \label{eq:hamQ}
\end{align}
Applying the argument developed by Shastry and Sutherland \cite{shastry1981,sutherland1983}, we can prove the quantum dimer nature of the ground state for part of the phase diagram. First, the diagonalization of Hamiltonian (\ref{eq:hamQ}) for a single triangle gives a lowest energy of $e_0 = -\frac{3J_2}{8}$ for $J_2\geq 2$ and $e_0 = \frac{J_2}{8} - 1$ for $J_2\leq 2$. Then, we consider the singlet state $\ket{\psi_0} = \prod_{\langle \langle i, j \rangle \rangle} \frac{\ket{\uparrow_i\downarrow_j} - \ket{\downarrow_i\uparrow_j}}{\sqrt{2}}$ for all pairs of spins on a $J_2$ link, whose energy per site is $-\frac{3J_2}{8}$. Since there are as many spins as triangles on our 3D SS lattice, the variational principle automatically implies that the singlet state is ground state for $J_2\geq 2$ \cite{shastry1981,sutherland1983}. Hamiltonian  (\ref{eq:hamQ}) thus offers a relatively rare example of an exact solution for a quantum spin model in three dimensions.

\begin{figure}[t]
    \centering
    \includegraphics[width=\linewidth]{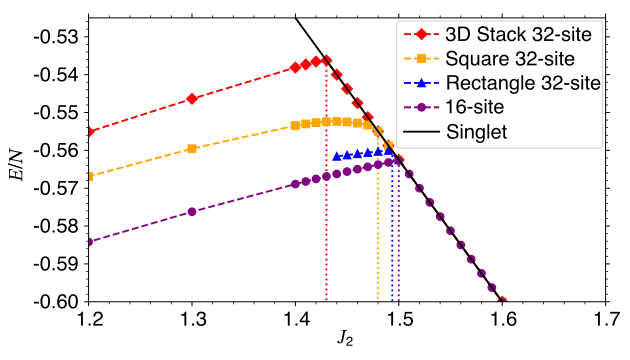}
    \caption{\justifying\small
    \textit{Quantum $S=1/2$:} Ground-state energy obtained from Exact Diagonalization for four different spin clusters with periodic boundary conditions: a 2D square of 16 (violet) and 32 (orange) sites, a 2D rectangle of 32 sites ($4\times 8$, blue) and a 3D cuboid of 32 sites made of two 16-site cubic unit cell on top of each other along the $z-$axis. The black line is the dimer-singlet energy and the dashed lines are simple guide-to-the-eye connecting data points.
    }
    \label{fig:ED}
\end{figure}

This solution is, however, only valid for $J_2\geq 2$. The rest of the phase diagram requires complementary approaches. In 2D, the dimer singlet phase persists down to $J_2=1.48$ \cite{Koga2000b,Corboz2013}, making space for a square-plaquette phase for $1.28 < J_2 < 1.48$ followed by a quantum spin liquid candidate for $1.22 < J_2 < 1.28$ and finally the N\'eel state for lower values of $J_2$ \cite{yang2022,viteritti2025,nutakki2025}.

In 3D, while the N\'eel state is also expected for small $J_2$, the existence of intermediate phases is an open question. The square-plaquette phase is actually incompatible with the 3D geometry whose smallest loop is made of seven sites. While valence bond solids of longer loops are possible, they are expected to be less stable. Hence, whether a different plaquette phase, either with larger resonating loops or on squares with $J_2$ bonds, remains a ground state in 3D, or whether it is replaced by another phase yet to be discovered, one can intuitively expect the dimer singlet to remain stable down to lower $J_2$ values. We tested the idea with Exact Diagonalization up to $N=32$ sites in Fig.~\ref{fig:ED}. A 16-site cluster with periodic boundary conditions is too small, and can be seen either as a square unit cell in 2D or a cubic unit cell in 3D. Hence, two 16-site clusters next to each other in the $x-$ or $y-$direction cannot really distinguish between 2 and 3 dimensions (blue triangles); but placing them on top of each other along the $z-$axis is only possible in 3D (red diamonds). Hence the difference between the blue and red data of Fig.~\ref{fig:ED} is the minimal difference between the original 2D and new 3D SS models. ED results are indeed consistent with a more stable dimer phase in 3D, even if further works are necessary to confirm it.

\section{Discussion}

The Shastry-Sutherland lattice is a cornerstone of frustrated magnetism. By extending its structure to three dimensions while keeping its connectivity intact, the present work opens the possibility to explore its exotic properties under the scope of dimensional enlargement. The stability of the up-up-down highly degenerate phase at finite field is an interesting aspect of the classical models, which is expected to remain stable over a broad range of parameters at finite temperature \cite{Moliner2009}. But the most interesting future direction is probably the quantum one.

The zero-field quantum phase diagram would be the first question to address. While our results suggest a more stable dimer singlet phase, the fate of the 2D intermediate phases is unknown, especially since the square-plaquette is impossible in 3D. In particular, is there a quantum spin liquid or a deconfined phase transition ? And what about magnetization plateaux in a field ? Since plateaux usually have large magnetic unit cell, one can a priori expect distinct behaviors between two and three dimensions. Also, since the 3D lattice is non-centrosymmetric, Dzyaloshinskii-Moriya interactions are allowed by symmetry; in that context does the topological nature of triplon bands \cite{romhanyi2015} persist in 3D or not ?

Finally, even though the present work is purely theoretical, the lattice of Fig.~\ref{fig:latt} is not completely unrealistic; it is, after all, a tetragonal deformation of the well-known pyrochlore lattice with breathing anisotropy. In that sense, realistic models would probably need to include a weak $J_3$ coupling on the ``top'' pyrochlore bond that was taken out in our lattice construction. This motivates considering the parameter path from the pyrochlore to the 3D Shastry-Sutherland lattice.

\textbf{Acknowledgments:} The authors would like to thank Frederic Mila for useful discussions. K.S-S. is supported by the program QuanTEdu-France. K.S.-S. and L.J. acknowledge financial support from Grant No. ANR-23-CE30-0038-01. Simulations were performed on the ``Mésocentre de Calcul Intensif Aquitain'' (MCIA).

\bibliographystyle{apsrev4-2}
\bibliography{biblio}

\end{document}